\documentclass[12pt]{article}
\batchmode

\begin{document}

\date{}

\title{A Simple Approximation for a Hard Routing Problem}

\author{
\vspace*{3mm}
\Large Rupei Xu \hspace*{7mm} Andr\'as Farag\'o \\
Department of Computer Science \\
The University of Texas at Dallas \\
Richardson, Texas\\
}

\maketitle
       \thispagestyle{empty}

    \noindent
    {\bf Abstract {\it
We consider a routing problem which plays an important role in several applications, 
primarily in  communication network planning and VLSI layout design. 
The original underlying graph algorithmic task  is called 
{\em Disjoint   Connecting   Paths}   problem. In most applications one can encounter its capacitated
generalization, which is known as the {\em Unsplittable Flow}  problem. 
These algorithmic tasks  are very  hard   in  general,   but
various efficient (polynomial-time)
approximate solutions  are known. Nevertheless, the approximations tend
to be rather complicated, often rendering them impractical in large, complex networks.
Our  goal is  to present  a solution  that
provides a simple,
efficient  algorithm  for  the  unsplittable  flow problem in large directed
graphs. The simplicity is achieved by sacrificing a small part of  the
solution space. This also represents a novel paradigm of approximation:
rather than giving up finding an exact solution, we restrict the solution space to its 
most important subset and exclude those that are marginal in some sense.
Then we find the {\em exact} optimum efficiently within the subset.
Specifically, the sacrificed parts (i.e., the marginal instances) only contain scenarios where some
edges  are  very  close  to  saturation.  Therefore,  the  excluded part is not
significant,  since  the excluded almost  saturated  solutions  are  typically
undesired in practical applications, anyway.


    }
        }

\newpage
\section{Introduction}
\label{dispath}

The  {\em  Disjoint  Connecting  Paths  problem}  is  the
following decision task:

{\bf Input:}
a set of node pairs $(s_1,t_1),\ldots, (s_k,t_k)$ 
in a graph. 

{\bf Task:}
Find edge disjoint paths
$P_1,\ldots,P_k$, such that  $P_i$ connects $s_i$  with $t_i$ for  each
$i$.

This is  one of the  classical {\bf NP}-complete problems  that appears
already at  the sources  of {\bf NP}-completess  theory, among  the original
problems of Karp \cite{karp}. It remains {\bf NP}-complete both for directed
and undirected  graphs, as  well as  for the  edge disjoint and vertex
disjoint  paths  version.   The  corresponding  natural   optimization
problem, when  we are  looking for  the maximum  number of  terminator
pairs that can be connected by disjoint paths is {\bf NP}-hard.

There is also a capacitated  version of the Disjoint Paths
Problem, also known  as the {\em  Unsplitting Flow problem}.  In this
task   a   {\em  flow   demand}   value   is   given   for   each
origin-destination pair $(s_i,t_i)$,  as well as  a capacity value  is
known for each  edge. The requirement  is to find  a system of  paths,
connecting the respective source-destination pairs, such that the {\em
capacity constraint}  of each  edge is  obeyed, i.e.,  the sum  of the
flows of  paths  that  traverse  the  edge  cannot be more than the
capacity of the edge. The name {\em  Unsplitting Flow} expresses 
the requirement that between each source-destination pair the flow must follow a single route, 
it cannot split.
 Note that here the disjointness of the paths themselves is
not  required  {\em  a  priori},  but  can be enforced by the capacity
constraints. The Unsplitting Flow problem is important in communication network design
and routing applications.

In this paper, after reviewing some existing results, we show that the
Unsplitting Flow problem,  which is {\bf NP}-complete,  becomes efficiently
solvable  by a relatively simple algorithm if  we  impose  a  mild
and practically well justifiable restriction on the instance.

\section{Previous Results}

Considerable work was done  on the Disjoint Paths Problem,
since its first appearance as an {\bf NP}-complete problem in \cite{karp} in
1972.

One direction of research deals with finding the ``heart" of the 
difficulty: which are the simplest restricted cases that still  remain
{\bf NP}-complete? (Or  {\bf NP}-hard if  the optimization  version is considered,
where we  look for  the maximum  number of  connecting paths, allowing
that possibly  not all  source-destination pairs  will be  connected).
Kramer and van Leeuwen \cite{kramer} proves, motivated by VLSI  layout
design,  that  the  problem  remains  {\bf NP}-complete  even  for graphs as
regular  as  a  two  dimensional  mesh.  If  we  restrict ourselves to
undirected planar graphs with each vertex having degree at most three,
the  problem  also  remains  {\bf NP}-complete,  as proven by Middendorf and
Pfeiffer \cite{middendorf}. The  optimization version remains  {\bf NP}-hard
for trees with parallel edges, although there the decision problem  is
already solvable in polynomial time \cite{garg}.

The restriction  that we  only allow  paths which  connect each source
node with a  {\em dedicated} target  is essential. If  this is relaxed
and we are satisfied with edge disjoint paths that connect each source
$s_i$ with {\em some} of  destinations $t_j$ but not necessarily  with
$t_i,$ then the problem  becomes solvable with classical  network flow
techniques. Thus, the prescribed matching of sources and  destinations
causes a dramatic change in the problem complexity. Interestingly,  it
becomes already {\bf NP}-complete if we  require that just {\em one}  of the
sources is connected to a  dedicated destination, the rest is  relaxed
as above (Farag\'o \cite{fapd}).

Another  group  of   results  produces  polynomial   time  algorithmic
solutions for finding the paths, possibly using randomization, in {\em
special  classes}  of  graphs.  For  example,  Middendorf and Pfeiffer
\cite{middendorf}  proves   the  following.   Let  us   represent  the
terminator pairs  by {\em  demand edges}.  These are  additional edges
that connect a source with its destination. If this extended graph  is
embeddable in the plane  such that the demand  edges lie in a  bounded
number of faces of the original graph, then the problem is solvable in
polynomial time.  (The faces  are the  planar regions  bordered by the
curves that  represent the  edges in  the planar  embedding, i.e.,  in
drawing the  graph in  the plane).  Thus, this  special case  requires
that, beyond the planarity of the extended graph, the terminators  are
concentrated in a constant number of regions (independent of the graph
size), rather than spreading over the graph.

A   deep   theoretical   result,   due   to   Robertson   and  Seymour
\cite{robertson}, is that  for general graphs  the problem
can be  solved in  polynomial time  if the  number $k$  of paths to be
found  is  {\em  constant}  (i.e.\  cannot  grow  with the size of the
graph). Broder, Frieze, Suen and Upfal \cite{broder} consider the case
of {\em random graphs} and provide a randomized algorithm that,  under
some technical conditions, finds  a solution with high  probability in
time $O(nm^2)$ for a graph of $n$ vertices and $m$ edges.

Another  line  of  research  aims  at  finding  approximations  to the
optimization   version.    An   algorithm    is   said    to   be   an
$f(n)$-approximation  if  it  can  connect  a subset of the terminator
pairs by disjoint paths such that this subset is at most $f(n)$  times
smaller than the optimum in a  graph of $n$ vertices. For example,  in
this terminology a 2-approximation  algorithm always reaches at  least
the half of  the optimum, or  an $O(\log n)$-approximation  reaches at
least a  $c/\log n$  fraction of  the optimum,  for $n>n_0$  with some
constants $c, n_0.$

Various  approximations  have  been  presented  in the literature. For
example,  Garg,   Vazirani  and   Yannakakis  \cite{garg}   provide  a
2-approximation  for  trees  with  parallel  edges.  Aumann and Rabani
\cite{aumann} gives an $O(\log n)$-approximation for the 2-dimensional
mesh.  Kleinberg  and  Tardos  \cite{kleinberg}  present  an   $O(\log
n)$-approximation for a  larger subclass of  planar graphs, they  call
``nearly Eulerian, uniformly high-diameter planar graphs" (the  rather
technical  definition  is  omitted  here).  For  the  general  case an
approximation factor of $\min\{\sqrt{m}, m/opt\}=O(\sqrt{m})$ is known
to  be  achievable  (Srinivasan  \cite{srinivasan}),  where $m$ is the
number of edges and $opt$ is the optimum, i.e., the maximum number  of
disjoint  connecting  paths  between  the  source-destination   pairs.
Similar  bounds  apply  for  the  Unsplitting  Flow problem, as well.
Bounds have been also found  in terms of special (less  trivial) graph
parameters. For  example, Kolman  and Scheideler  \cite{kolman} proves
that an  efficient $O(F)$  approximation exists,  where $F$  is the so
called {\em flow number} of the graph. Although the flow number can be
computed  in  polynomial  time  \cite{kolman},  it  is an indirect
characterization of the graph.

\section{A Simple Practical Approximation}

The various above referenced  solutions are rather complicated,  which
is certainly not helpful for practical applications, in particular in large, complex networks. 
Our approach  for providing a simple solution to the unsplitting flow problem based  on
the following idea. We ``cut down" a small part of the solution  space
by slightly reducing the edge  capacities. In other words, we  exclude
solutions that are close to saturating some edge, as explained below.

Let $V_i$ be the given flow demand of the $i^{\rm th}$  connecting
path. We normalize these demands such that $V_i\leq 1$ for every  $i$.
Let $C_j$ be the capacity of  edge $j.$ The graph is assumed  directed
and the edges are numbered from 1 through $m$. Recall that a  feasible
solution of the problem is  a set of connecting (directed)  paths that
satisfy the edge capacity constraints,  that is, on each edge  $j$ the
sum of the $V_i$ values of those paths that traverse the edge does not
exceed  $C_j.$  As  mentioned  earlier,  deciding  whether  a feasible
solution exist at all is a difficult ({\bf NP}-complete) problem.

On the other  hand, not all  feasible solutions are  equally good from
the practical viewpoint. For example,  if a route system in  a network
saturates or nearly saturates some  links, then it is not  preferable
because  it  is  close  to  being  overloaded. For this reason, let us
assign a parameter $0<\rho_j<1$ to  each edge $j,$ such that  $\rho_j$
will act as a  ``safety margin" for the  edge. More precisely, let  us
call  a  feasible  solution  a  {\em  safe  solution}  with parameters
$\rho_j,\; j=1,\ldots,m,$ where $m$ is the number of edges, if it uses
at most $\widetilde{C}_j=\rho_j C_j$ capacity on edge $j.$

Now, the interesting  thing is that  if we restrict  ourselves to only
those  cases  when  a  safe  solution   exists,  then  the   hard
algorithmic problem becomes solvable by a relatively simple randomized
algorithm. With very high probability the algorithm finds a  solution
in polynomial time, whenever there exists  a safe solution. 

 The price is  that we exclude
those cases when a {\em feasible} solution still possibly exists, but  there
is no {\em safe} solution. This means, in these cases all feasible solutions 
are undesirable, in the sense that they make some edges nearly saturated.
In these marginal cases the algorithm may find no
solution at all. This approach constitutes a new avenue to approximation, in the 
sense that instead of giving up finding an exact solution, we rather restrict 
the search space to a (slightly) smaller one. When, however, the algorithm finds 
any solution, then it is an {\em exact} (not just approximate) solution.

Now let us choose the safety margin $\rho_j$
for a graph of $m$ edges as
\begin{equation} \label{safmargin}
\rho_j=1-(e-1)\,\sqrt{\frac{{\rm ln}\, 2m}{C_j}}\approx
1-1.71\,\sqrt{\frac{{\rm ln}\, 2m}{C_j}}
\end{equation}
where ln denotes the natural logarithm $\log_e$.
Note that $\rho_j$
tends to 1 with growing $C_j$, even if
the graph also grows, but $C_j$ grows faster than the logarithm of
the graph size, which is very reasonable (note that doubling the number of
edges will increase the natural logarithm by less than 1).
For example, if in a graph each
edge capacity is 1000 units, measured in relative units, such that the
maximum path flow demand is 1, and  the graph  has 200 edges, then
$\rho\approx 0.97.$

Now we  outline how  the algorithm  works. To  make it  even closer to
practical applications, we also assume that cost factors are  assigned
to the edges  and we are  looking for a  feasible solution with  small
cost, where  the cost  incurred on  an edge  is proportional  with the
demand routed through it.

\medskip
\noindent
{\bf Algorithm}
\begin{description}

\item[{\it   Step   1\,\,\,}]   {\em   Initialization}\\  Compute  the
$\widetilde{C}_j=\rho_j  C_j$  values  with  $\rho_j$ set according to
(\ref{safmargin}).

\item[{\it  Step  2\,\,\,}]  {\em   Flow  relaxation}  \\  Solve   the
continuous minimum cost multicommodity flow relaxation of the problem,
using the $\widetilde{C}_j$ capacities\footnote{
Note that although this phase finds a flow of the
required value between  each source-destination pair,  
it does not yet
provide the required {\em unsplittable} flow, since the found flow typically branches
arbitrarily into
small parts rather than going on one path, this is why it is called  a
relaxation of the problem.}.
This can be done  by standard
linear programming.
In  case the flow problem has  no solution
then declare ``no safe solution exists" and STOP.

\item[{\it Step 3\,\,\,}] {\em Randomized Rounding via Random Walk} \\
For  each  source-destination  pair  $u_i,v_i$  find  a  path  via the
following randomized rounding procedure. Start at the source and  take
the  next  node  such  that  it  is drawn randomly among the successor
neighbors  of  the  source,  with  probabilities  proportional  to the
$i^{\rm th}$  commodity flow  values on  the edges  from $u_i$  to the
successor neighbors in the directed graph. Continue this in a  similar
way: at each  node choose the  next one among  its successor neighbors
randomly, with probabilities that are proportional to the $i^{\rm th}$
commodity flow values.  Finally, upon arrival  at $v_i,$ we  store the
found $(u_i,v_i)$ path.

\item[{\it Step  4\,\,\,}] {\em  Feasibility Check  and Repetition} \\
Having found a system of  paths in the previous steps,  check whether
it is a  feasible solution. If  so, then STOP,  else repeat from  {\em
Step 2.} \\ If  after repeating $r$ times  ($r$ is a fixed  parameter)
none of the runs are  successful then declare ``No solution  is found"
and STOP.

\end{description}

It is clear from the above informal description that the algorithm has
practically feasible complexity, since the most complex part of it  is
solving  a  multicommodity  flow  problem  that  can be done by linear
programming. It is repeated $r$ times where $r$ is a parameter, chosen
by us. The main  property of the algorithm  is shown in the  following
theorem.

\medskip\noindent
{\bf Theorem 1}
{\em If a  safe solution  exists, the  algorithm finds  a feasible
solution with probability at least $1-2^{-r}.$}

\medskip\noindent
{\bf  Proof.}  Since  a  safe  solution is also a
feasible solution of the multicommodity flow relaxation, therefore, if
there is no flow  solution in {\em Step  2,} then no safe  solution can exist
either.

{\em Step 3} transforms the flow solution into paths. To see that they
are indeed paths, observe that looping cannot occur in the  randomized
branching procedure, because if a circle arises on the way, that would
mean a  circle with  all positive  flow values  for a given commodity,
which  could  be  canceled  from  the  flow  of  that commodity, thus
contradicting to the minimum cost property of the flow. Furthermore, since looping 
cannot occur, we must reach the destination via the procedure in at most 
$n$ steps, where $n$ is the number of nodes.

Now a key observation is that if we build the paths with the described
randomization between the  $i^{\rm th}$ source-destination  pair, then
the {\em expected value} of the load that is put on any given edge  by
these paths will  be exactly the  value of the  $i^{\rm th}$ commodity
flow  on  the  link.  This  follows  from  the fact that the branching
probabilities are flow-proportional.

From the above we know that the total expected load of an edge,
arising
form the randomly chosen  paths, is equal to  the total flow value  on
the edge. What we have to  bound is the deviation of the  {\em actual}
load from this expected value. Let $F_j$ be the flow (=expected  load)
on edge $j.$ This  arises in  the randomized  procedure as $$F_j={\rm
E}\Big(\sum_i V_i X_i\Big) ,$$ where  $X_i$ is a random variable  that
takes the value 1 if  the $i^{\rm th}$ path contributes to the edge
load, otherwise
it is 0. The construction implies that these random variables are
independent.

Now  consider  the  random  variable  $$\Psi_j=\sum_i V_i X_i.$$ We have
${\rm E}(\Psi_j)=F_j.$  The probability  that $\Psi_j$  deviates form  its
expected value by more than a factor of $\delta$ can be bounded by the
tail inequality found in \cite{raghavan}:
$$\Pr \Big(\Psi_j>(1+\delta) F_j\Big)<
\left(\frac{e^\delta}{(1+\delta)^{(1+\delta)}}\right)^{F_j}.$$
It can be calculated from this \cite{raghavan} that
if we want to guarantee that the bound does not exceed a given value
$\epsilon>0,$ then it is sufficient to satisfy
$$\delta \leq (e-1)\,\sqrt{\frac{{\rm ln\,}(1/\epsilon)}{F_j}}.$$
Let us choose $\epsilon=1/(2m).$ Then we have
$$\Pr \left(\Psi_j>\Big(1+(e-1)\,\sqrt{({\rm ln\,}2m)/F_j}
\Big) F_j \right) $$
$$< \frac{1}{2m}.$$
Since the bound that we do not want to exceed is the edge capacity
$C_j,$ therefore, if
\begin{equation} \label{cj}
C_j\geq\Big(1+(e-1)\,\sqrt{({\rm ln\,}2m)/F_j} \Big) F_j
\end{equation}
is satisfied, then we have
$$\Pr (\Psi_j>C_j) < \frac{1}{2m}.$$
If this holds for all edges, that yields
\begin{eqnarray}    \nonumber
\Pr (\exists j:\;
\Psi_j>C_j) & \leq &
\sum_{j=1}^m \Pr (\Psi_j>C_j) \\ \nonumber
& < & m\frac{1}{2m}  \\ \nonumber
& = & \frac{1}{2}.
\end{eqnarray}
Thus, the probability that the found path system is not feasible is
less than $1/2.$ Repeating the procedure $r$ times with independent randomness, the
probability that none of the trials  provide a feasible solution
is bounded by $1/2^r,$ that is, the failure probability
becomes very small, already for moderate values of $r.$

Finally, expressing $F_j$ form (\ref{cj}) we obtain
$$F_j\leq
C_j\left(1-(e-1)\,\sqrt{\frac{{\rm ln}\, 2m}{C_j}}\right)
=\rho_j C_j,$$
which shows that the safety margin is correctly chosen, thus
completing the proof.

\hfill $\spadesuit$

\section{Conclusion}

We have presented  a simple, efficient  solution for the {\bf NP}-complete Unsplittable
Flow  problem  in  directed  graphs.  The  simplicity and efficiency  is  achieved by
sacrificing a small  part of the  solution space. The  sacrificed part
only contains scenarios where some edges are very close to saturation.
Therefore, the  loss is  not significant,  since the  almost saturated
solutions are typically undesired  in practical applications, such  as
network design. 

The approach constitutes a new avenue to approximation, in the 
sense that instead of giving up finding an exact solution, we rather restrict 
the search space to a (slightly) smaller one. When, however, the algorithm finds 
any solution, which happens with high probability, then it is an {\em exact} 
(not just approximate) solution. In this paper we only laid down the theoretical foundations
of the approach, numerical validation is planned for a later paper.

\end{document}